\documentclass[%
 reprint,
 amsmath,amssymb,
 aps,
]{revtex4-2}

\usepackage[normalem]{ulem}
\usepackage{graphicx}
\usepackage{dcolumn}
\usepackage{bm}

\usepackage{amsthm}
\usepackage{float}
\usepackage{adjustbox}
\usepackage{ulem}
\usepackage{mathrsfs}
\usepackage{xcolor}
\usepackage{subcaption}
\usepackage{mwe}

\usepackage{algorithm,algcompatible,algorithmicx}
\usepackage{algpseudocode}

\algnewcommand\algorithmicinput{\textbf{Input:}}
\algnewcommand\INPUT{\item[\algorithmicinput]}
\algnewcommand\algorithmicoutput{\textbf{Output:}}
\algnewcommand\OUTPUT{\item[\algorithmicoutput]}
\algnewcommand\algorithmicoptional{\textbf{Optional:}}
\algnewcommand\OPTIONAL{\item[\algorithmicoptional]}

\begin{document}

\title{Optimization of passive superconductors for shaping stellarator magnetic fields}%

\author{Alan A. Kaptanoglu}
 \email{alankaptanoglu@nyu.edu}
 \affiliation{Courant Institute of Mathematical Sciences, New York University, New York, NY, 10012, USA      \looseness=-1}
 \affiliation{Department of Mechanical Engineering, University of Washington,
Seattle, WA, 98195, USA \looseness=-1
}

\author{Matt Landreman}
\affiliation{IREAP, University of Maryland, College Park, MD, 20742, USA}

\author{Michael C. Zarnstorff}
\affiliation{Princeton Plasma Physics Laboratory, P.O. Box 451, Princeton, New Jersey 08543, USA}
\begin{abstract}
We consider the novel problem of optimizing a large set of passive superconducting coils (PSCs) with currents induced by a background magnetic field rather than power supplies. In the nuclear fusion literature, such coils have been proposed to partially produce the 3D magnetic fields for stellarators and provide passive stabilization. We perform the first optimizations of PSC arrays with respect to the orientation, shape, and location of each coil, jointly minimized with the background fields. We conclude by generating passive coil array solutions for four stellarators. 
 \end{abstract}
\maketitle

\section{Introduction}
\label{sec:intro}
The goal of inverse magnetostatic optimization is to find a configuration of external magnets to produce a given magnetic field in a volume or on a surface. 
Inverse magnetostatics is an ill-posed inverse problem because many different magnet designs can produce a nearly identical target magnetic field via the Biot-Savart law. This important problem appears across the scientific literature, including for magnetic resonance imaging (MRI)~\cite{hidalgo2010theory, chen2017electromagnetic, cooley2017design, ren2018design}, beam optics in particle accelerators~\cite{russenschuck2011field}, permanent magnet motors~\cite{gieras2009permanent}, and degaussing magnetized objects~\cite{bekers2013degaussing}.

In addition, it is a critical component of designing stellarators, a class of plasma and nuclear fusion experiments with 3D magnetic fields. 
%
%
After prescribing a desired equilibrium plasma, a set of magnets is determined that can produce a magnetic field that matches the boundary conditions at the assumed plasma surface, subject to engineering constraints such as minimum coil-to-coil distances, maximum forces on the coils~\cite{robin2022minimization}, maximum curvature on the coils~\cite{zhu2017new}, and many other requirements~\cite{imbert2024introduction}. 
Unfortunately, stellarator coils are typically complex three-dimensional shapes, often causing high costs and imposing challenges associated with manufacturing, assembly, and maintenance~\cite{erckmann1997w7,strykowsky2009engineering}. 

There are some alternative coil options for producing the required magnetic fields. 
For instance, permanent magnets can be cost-effectively used with modular coils to design university-scale devices~\cite{qian2021stellarator,qian2022simpler,qian2023design,hammond2020geometric,kaptanoglu2022greedy,kaptanoglu2022permanent,hammond2024improved,guodong2024}, but tens of thousands of magnets are typically required, the magnets have limited field strengths, and they demagnetize in the high-field environments relevant for nuclear fusion reactors. For reactor-scale designs, start-up companies~\cite{gates2023thea, zarnstorff2023stellarex} have proposed large arrays of relatively small saddle/window-pane coils (coils that are not topologically linked with the plasma) to provide the benefits of permanent magnets (simplicity, modularity, cost, etc.) without the primary downside (limited field strengths, cannot demagnetize). We will refer to these designs as dipole arrays because these small coils primarily produce magnetic dipole fields. Unlike permanent magnets, each of these coils still requires power supplies and previous work found that coil-coil forces can easily become a limiting issue~\cite{kaptanoglu2024reactorscalestellaratorsforcetorque}. 

Superconducting coils can also \textit{passively} produce a field in response to a background magnetic field, eliminating the need for power supplies and potentially simplifying coil design and assembly. Like the permanent magnets or dipole arrays, arrays of passive superconducting coils (PSCs) cannot produce a net toroidal magnetic flux~\cite{helander2020stellarators}. It follows that a set of toroidal field (TF) coils is still needed; the goal of the array in fusion devices is to simplify the cost and design of these TF coils. Importantly, the TF coils additionally provide the required background magnetic field.

In this work, we present the first large-scale joint optimizations of PSC arrays and the background field. While we focus primarily on optimizing a set of identical, planar, circular passive coils for stellarators, we illustrate that the methods in this work can be used for arbitrarily shaped passive coils and arbitrary inverse magnetostatic problems. The entirety of this work is implemented and can be reproduced in the open-source SIMSOPT code~\cite{landreman2021simsopt}. 

\subsection{Passive superconducting coils}
Consider two sets of coils, an array of PSCs and a set of coils with power supplies that can produce a background field. Before the latter
coils are energized, the PSCs are lowered below
the critical temperature for superconductivity. The passive coils would enclose zero
net magnetic flux and, when the powered coils are energized, currents
are induced in the passive coils to maintain zero enclosed flux. 
During the initial energizing of the TF coils, large electric fields can be generated between all the PSCs and TF coils through Faraday's law. Nonetheless, this will be a transient effect and the electric fields will decay away. It is in this magnetostatic limit that the magnetic field shaping is important, and subsequently it is the regime for which we will perform optimization. 
Although we are unsure of a scientific motivation, we note that in a variation of this approach the passive coils could be made to
have a nonzero flux, by imposing a specified background field while they are lowered them below the critical temperature.

The plasma is also a conductor of currents, meaning that there is coupling to all the coils. This coupling is always ignored in stellarator optimization calculations. If the coils are connected to power supplies, coupling over long timescales can be removed by simple algorithms such as a proportional-integral-derivative (PID) controller that restores the prescribed current waveforms. In contrast, the long timescale coupling between passive coil currents and plasma currents cannot be removed. Fast timescale coupling, e.g. when the plasma experiences a sudden disruption, is a problem for any superconducting coil.
An investigation of passive stabilization or destabilization would require time-dependent simulations not pursued in the present work.

Passive superconducting tiles, also referred to as superconducting pucks or superconducting monoliths, are a variation of passive coils where solid superconducting blocks are used instead of coil loops. These tile arrays were briefly considered in Bromberg et al.~\cite{bromberg2011stellarator} to prescribe an exact fixed-boundary magnetic flux surface. They also considered passive tiles for magnetic field shaping but numerical optimization was not pursued. 
The optimization of such tiles would be essentially the same as proposed here.

\section{Superconductor magnetostatics}
The goal of this work is to construct a formulation of magnetostatic optimization for PSC arrays.
There are two distinct properties of superconductors to consider here.
First, superconductors have zero resistance, i.e. flux-freezing. From zero resistance, it follows
that the line integral of the electric field around a superconducting
loop is zero. Hence by Faraday's law, the magnetic flux through such
a loop does not change in time. 
Assume that a circular PSC starts with zero net flux through the loop $C_i$, and then a background vector potential $\bm A_0$ is initialized.
A current $I_i$ is induced to cancel the flux $\psi_i$ from the background field,
\begin{align}
\label{eq:currents_exact}
    I_i = \frac{\psi_i}{L} =  \frac{1}{L}\int_{C_i}\bm A_0(\bm \gamma_i)\cdot d\bm l_i,
\end{align}
where $\bm \gamma_i$ are the points along the curve and $d\bm l_i$ is the differential arc length along the curve. The self-inductance $L$ can be efficiently computed for wires with circular or rectangular cross-sections~\cite{landreman2023efficient,hurwitz2024efficient}.
%
Note that this is an inductive current so it scales inversely with the number of wire turns.
It follows that the magnetic field from this coil will be independent of the number of turns so we drop this prefactor for the remainder of the work. Multiple turns can be used to reduce forces and stay below the critical current density in the superconductor. 
To estimate the size of the induced currents, consider a uniform background field $\bm B_0$ and a circularly shaped coil with major radius $R$ and circular cross section of minor radius $a$. Then some plausible values for a reactor scale design are: $R \sim 0.5$ m, $\bm B_0 \sim 5$ T, and $a / R \sim 0.1$, we get $I \sim 2.4$ MA from Eq.~\eqref{eq:currents_exact}. Larger currents can be driven in energized dipole arrays than in PSC arrays, but there is a tradeoff between the larger magnetic fields and the subsequently large coil-coil forces. For instance, minimizing forces during PSC array optimization is not typically required for the coils to stay within the material tolerances.
%
%
Lastly, the second distinct superconductor property is the Meissner effect, which we ignore for the purposes of the present work and justify this omission in Appendix~\ref{sec:meissner_discussion}.

\subsection{Determining the currents in a PSC array}\label{sec:forward_problem}
The currents induced in an array of PSCs can be found similarly to Eq.~\eqref{eq:currents_exact}.
Let $N$ be the number of passive coils considered in a given configuration. 
The currents in each of these PSCs represent $N$ unknowns and the condition that zero magnetic flux is maintained through the PSCs after energizing the background field coils represents $N$ constraints. 
%
Let $\mathbf{I}\in\mathbb{R}^N$ be a column vector of the currents $I_{j}$ in each
coil $j$. 
Let $\mathbf{L}\in\mathbb{R}^{N\times N}$ be the inductance matrix consisting of the mutual and self-inductances
$l_{ij}$. 
Let $\mathbf{\Psi}\in\mathbb{R}^N$
be a column vector consisting of the flux in each coil driven by the
energized TF coils. The condition that the total flux in each passive coil
vanishes is simply that the induced currents produce a magnetic flux that cancels the flux coming from the powered coils,
\begin{equation}
\label{eq:I_solve}
\mathbf{LI}+\mathbf{\Psi}= \bm 0.
\end{equation}
This is a moderate-size linear system of equations and
the $\mathbf{L}$ matrix is symmetric and positive-definite. Therefore $\bm I$ can be rapidly computed by performing a Cholesky decomposition of $\bm L$ followed by solving two triangular linear systems of equations.

\subsection{Optimizing PSC arrays}\label{sec:PSC_optimization_problem}
The goal of inverse magnetostatics is to find a set of coil shapes, locations, and currents that produce a desired magnetic field on a surface or in a volume. Eq.~\eqref{eq:I_solve} shows that the passive coil currents are entirely determined by the active coils and the geometry of the PSCs in the array, so the degrees of freedom to optimize are the active coil currents, shapes, and locations, as well as the PSC degrees of freedom, i.e. each PSC orientation and location. For geometric simplicity and potential mass production, coil shapes are primarily fixed to be planar circular coils in the present work, although this is not a requirement of the optimization problem. In Sections~\ref{sec:QA} and~\ref{sec:CSX} we optimize the coil shapes to achieve lower errors.

We formulate the following optimization problem using well-known coil objectives and new terms related to the coil forces and torques, following Kaptanoglu et al.~\cite{kaptanoglu2024reactorscalestellaratorsforcetorque}. Given a plasma surface $S$ and its normal vectors $\hat{\bm n}$, the primary objective for stellarator coil optimization is to minimize the surface-averaged magnitude of the $\bm B\cdot\hat{\bm n}$ errors,
\begin{align}
    K = \frac{1}{2}\int_S|\bm B\cdot\hat{\bm n}|^2dS.
\end{align}
For a normalized error, we will report $\langle \bm B \cdot\hat{\bm n}\rangle / \langle B \rangle$, where $\langle . \rangle$ defines an average over the plasma surface.
The TF coil curves are parametrized as finite Fourier series and the coefficients of these series become the degrees of freedom for optimization. The degrees of freedom for the PSCs are the normal vectors to the plane and their center point locations in space. Optimization of the geometrical shape of the PSCs has also been implemented although simple coil geometry is often an asset. We use the standard L-BFGS algorithm~\cite{liu1989limited} to solve the optimization problem.

Obtaining derivatives with respect to the PSC degrees of freedom is very challenging because each coil current depends on all the coils and their geometry. To deal with these complex dependencies and retain computational efficiency, we in part utilize the JAX package for autodifferentiation~\cite{bradbury2018jax} and in part rely on the parallelized C++ derivatives already implemented in SIMSOPT with respect to the various curve and magnetic field representations. For instance, for an objective such as $K$ that depends on the magnetic field, we compute the derivative $\partial_{\bm\eta} K$ with respect to the degrees of freedom $\bm \eta$ via,
\begin{align}
\frac{\partial K}{\partial \bm \eta} &= \frac{\partial K}{\partial \bm B_a} \left(\frac{\partial \bm B_a}{\partial \bm \gamma_a}\frac{\partial \bm \gamma_a}{\partial \bm \eta} + \frac{\partial \bm B_a}{\partial \bar{\bm \gamma}_a}\frac{\partial \bar{\bm \gamma}_a}{\partial \bm \eta} + \frac{\partial \bm B_a}{\partial \bm I_a}\frac{\partial \bm I_a}{\partial \bm \eta}\right) \\ \notag &+ \frac{\partial K}{\partial \bm B_p} \left(\frac{\partial \bm B_p}{\partial \bm \gamma_p}\frac{\partial \bm \gamma_p}{\partial \bm \eta} + \frac{\partial \bm B_p}{\partial \bar{\bm \gamma}_p}\frac{\partial \bar{\bm \gamma}_p}{\partial \bm \eta} + \frac{\partial \bm B_p}{\partial \bm I_p}\frac{\partial \bm I_p}{\partial \bm \eta}\right).
\end{align}
Here $\bm B_a$ is the active magnetic field, $\bm \gamma_a$ are points along the active coil curves, $\bar{\bm \gamma}_a$ are the tangent vectors along the active coil curves, $\bm I_a$ are the active coil currents, and the subscript $p$ denotes equivalent quantities for the PSCs. Since $\bm I_a$ is one of the degrees of freedom, $\partial_{\bm\eta}\bm I_a$ is identity or zero. In contrast, $\bm I_p$ is not a degree of freedom and must be determined through another chain of derivatives,
\begin{align}  \label{eq:Ip_solve}
&\frac{\partial \bm I_p}{\partial \bm \eta} = \left(\frac{\partial \bm I_p}{\partial \bm \gamma_p}\frac{\partial \bm \gamma_p}{\partial \bm \eta} + \frac{\partial \bm I_p}{\partial \bar{\bm \gamma}_p}\frac{\partial \bar{\bm \gamma}_p}{\partial \bm \eta}\right) \\ \notag &+ \frac{\partial \bm I_p}{\partial \bm A_a}\left(\frac{\partial \bm A_a}{\partial \bm \gamma_a}\frac{\partial \bm \gamma_a}{\partial \bm \eta} + \frac{\partial \bm A_a}{\partial \bar{\bm \gamma}_a}\frac{\partial \bar{\bm \gamma}_a}{\partial \bm \eta} + \frac{\partial \bm A_a}{\partial \bm I_a}\frac{\partial \bm I_a}{\partial \bm \eta}+\frac{\partial \bm A_a}{\partial \bm \gamma_p}\frac{\partial \bm \gamma_p}{\partial \bm \eta}\right).
\end{align}
Above $\bm A_a$ is the vector potential generated by the active coils.
Eq.~\eqref{eq:Ip_solve} reflects the fact that the PSC currents depend on all the curves and active currents. 
Note that Eq.~\eqref{eq:Ip_solve} also has a $\partial \bm A_a /\partial \bm \gamma_p$ term even when the active field coils are fixed during optimization. This is because the external field's vector potential is evaluated along each PSC curve, which is still changing during optimization. The accuracy of the Jacobian calculations were extensively verified against finite differences. 

\section{Results}\label{sec:results}
Stellarators are typically designed with two discrete symmetries: stellarator symmetry and $N_p$-field-period symmetry. From these symmetries, it follows that only the plasma shape and coils in the toroidal sector $[0, \pi / N_p)$ (a half-field-period) are designed, and the remaining plasma shape and coils are determined by enforcing the discrete symmetries. 
We present solutions for three reactor-scale and one university-scale stellarators, all of which are stellarator and field-period symmetric: the Landreman-Paul two-field period quasi-axisymmetric (QA) and four-field period quasi-helically (QH) symmetric designs~\cite{landreman2022magnetic}, the Schuett-Henneberg two-field period QA design~\cite{schuett2024exploring}, and the CSX design~\cite{baillod2024integrating}. 

The reactor-scale configurations are scaled to ARIES-CS parameters: average on-axis magnetic field strength $B_0 = 5.7$ Tesla and minor radius $r_0 = 1.7$m~\cite{najmabadi2008aries}. Tolerable coil forces are considered to be approximately $0.8$ MN/m based on reasonable estimates~\cite{hartwig2020viper,zhao2022structural,riva2023development,hartwig2023sparc,xu2025double}. For computing forces, we assume a plausible $100$ turns of wire for PSCs and $200$ turns of wire for the TF coils. 
The locations of the dipole coils are prescribed by uniformly initializing coils between an inner and outer toroidal surface defined by extending the plasma surface using the normal vector to the plasma surface times a fixed offset distance.
The inner surface is $1.5$m from the plasma or further because a large neutron absorbent blanket is required in a reactor-scale machine. PSCs that can intersect or interlink with other PSCs or the initial TF coils are simply removed before optimization starts. 
The minimum TF coil-coil distance is given as $0.8$m and we allow a minimum distance of half this value between the TF coils and the PSCs or between PSCs. As is standard in the literature, the accuracy of the final solution is verified with Poincaré plots and self-consistent equilibria calculations. More tightly packed configurations of passive coils or dipole coils are available if nonplanarity is relaxed, as in e.g. Hammond et al.~\cite{hammond2024framework}.

\subsection{Reactor-scale Landreman-Paul QA stellarator}\label{sec:QA}
\begin{figure}
    \centering
    \includegraphics[width=0.95\linewidth]{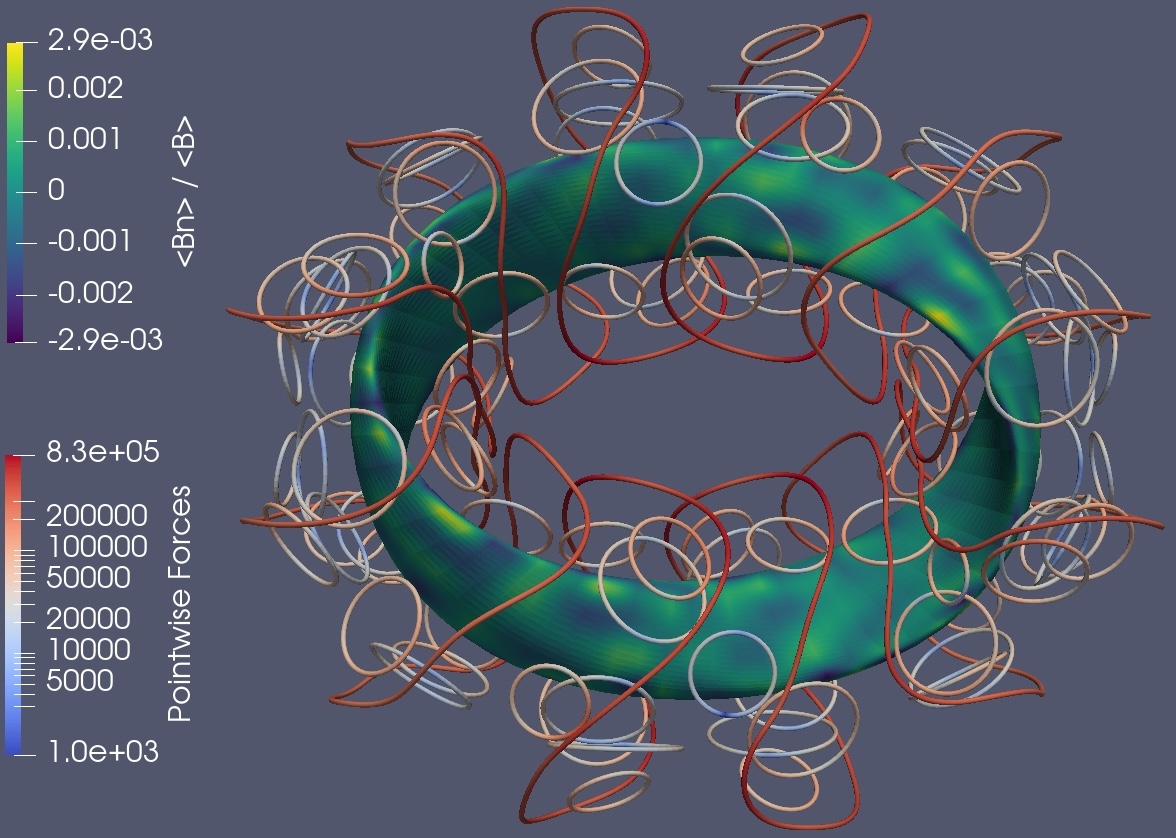}
    \caption{PSC array solution for the two-field-period Landreman-Paul QA stellarator.}
    \label{fig:QA}
    \includegraphics[width=0.95\linewidth]{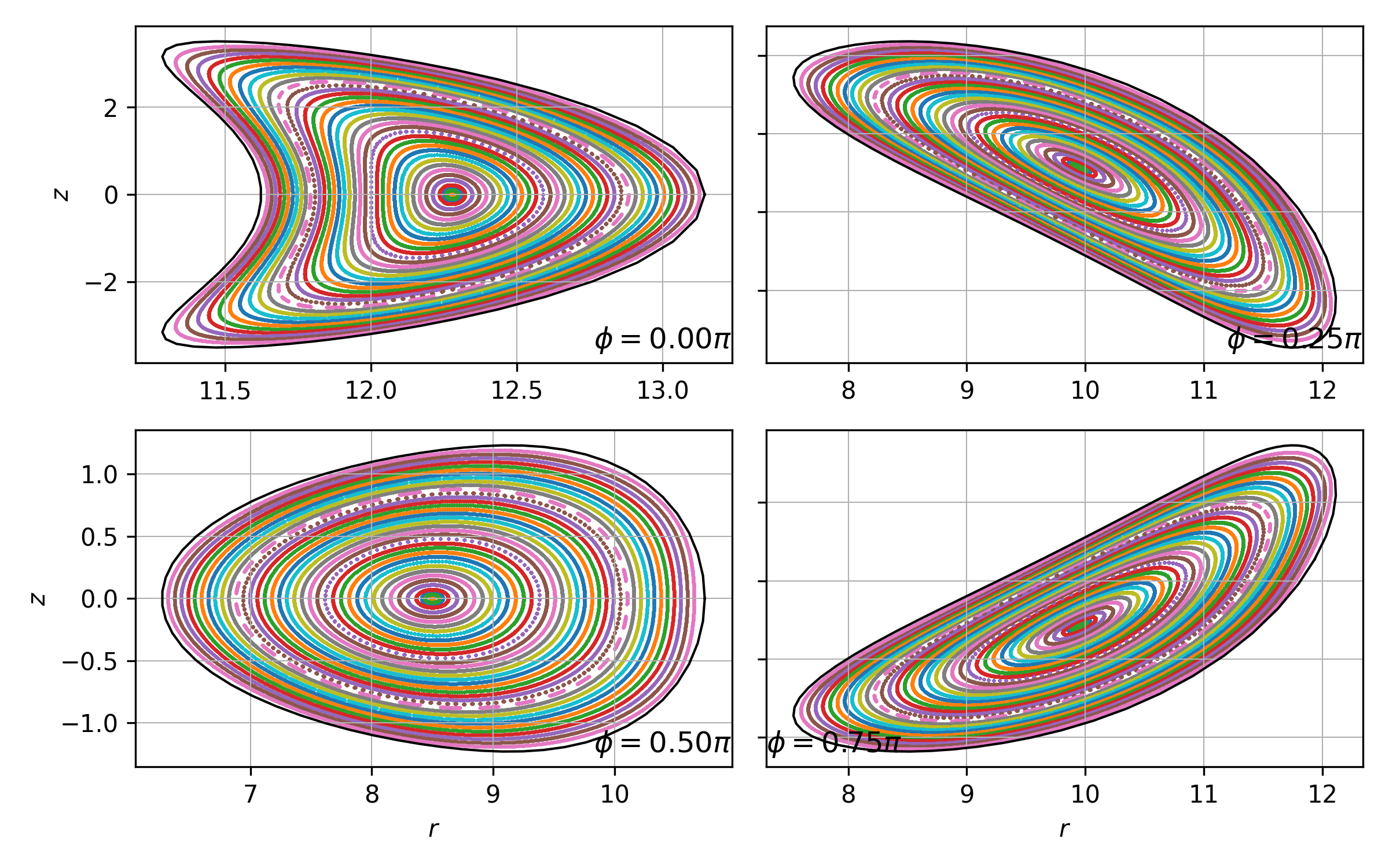}
    \caption{Poincaré plot for the Landreman-Paul QA coil design shows good magnetic surfaces.}
    \label{fig:QA_poincare}
\end{figure}

A PSC array design with three TF coils per half-field-period is shown in Fig.~\ref{fig:QA} for the two-field-period symmetric Landreman-Paul QA stellarator. Best results were found when the radii of the coils were allowed to vary in a last stage of optimization. The average normalized error of the final solution, $\langle\bm B\cdot\hat{\bm n}\rangle/\langle B\rangle \approx 5.9\times 10^{-4}$ is very low and produces good Poincaré plots in Fig.~\ref{fig:QA_poincare}. The average normalized two-term quasi-axisymmetry error,
\medmuskip=-1mu
\thinmuskip=-1mu
\thickmuskip=0mu
\begin{align}
    \sum_s\left\langle \left(\frac{(N - \iota M)\bm B \times \nabla B\cdot\nabla\psi - (MG + NI)\bm B \cdot\nabla B}{B^3}\right)^2\right\rangle,
\end{align}
\medmuskip=4mu
\thinmuskip=4mu
\thickmuskip=4mu
at approximately $7\times 10^{-5}$ is reduced by an order of magnitude from the original value of $7 \times 10^{-6}$. However, this solution still exhibits a very high degree of quasi-symmetry, substantially better than most designs in the literature. Above, $\iota$ is the rotational transform, $G(s)$ is proportional to the poloidal current outside the flux surface labeled by the normalized toroidal flux coordinate evaluated at $s = 0, 0.1, ..., 1$, $I(s)$ is proportional to the toroidal current inside the surface, $\psi$ is proportional to the toroidal flux, $\langle .\rangle$ brackets denote average over a flux surface, and $(M, N)$ are integers that specify the helicity of symmetry.

We use an array of twenty-five PSCs per half-field-period with final radii between $1.14-2.49$m and average radius $1.8$m. The optimized PSC coil currents vary between $0.51-2.9$ MA. Throughout these results, the reported currents are the total current across the 100 turns, so each turn carries $\sim 10$ kA. 
The TF coils have total length of $135$m, average coil curvature $\kappa \approx 0.55$m$^{-1}$, and forces within the material tolerances. The final minimum distance between TF coils is $1.51$m and between PSCs, or between a PSC and a TF coil, is $0.5$m. These TF coils are comparable in size, comparable in average curvature, and exhibit substantially larger TF coil-coil distances to the four TF coil solutions in Wechsung et al.~\cite{wechsung2022precise}. In addition, this solution is comparable to the dipole array solution in Kaptanoglu et al.~\cite{kaptanoglu2024reactorscalestellaratorsforcetorque}; the TF coils are necessarily longer for the PSC array (total length $135$m rather than $115$m) because the PSC currents are substantially smaller. However, in addition to lacking power supplies, PSC arrays have the benefit of easily keeping all the coil forces and torques tolerable. 


\subsection{Reactor-scale Landreman-Paul QH stellarator}
\begin{figure}
    \centering
\includegraphics[width=0.875\linewidth]{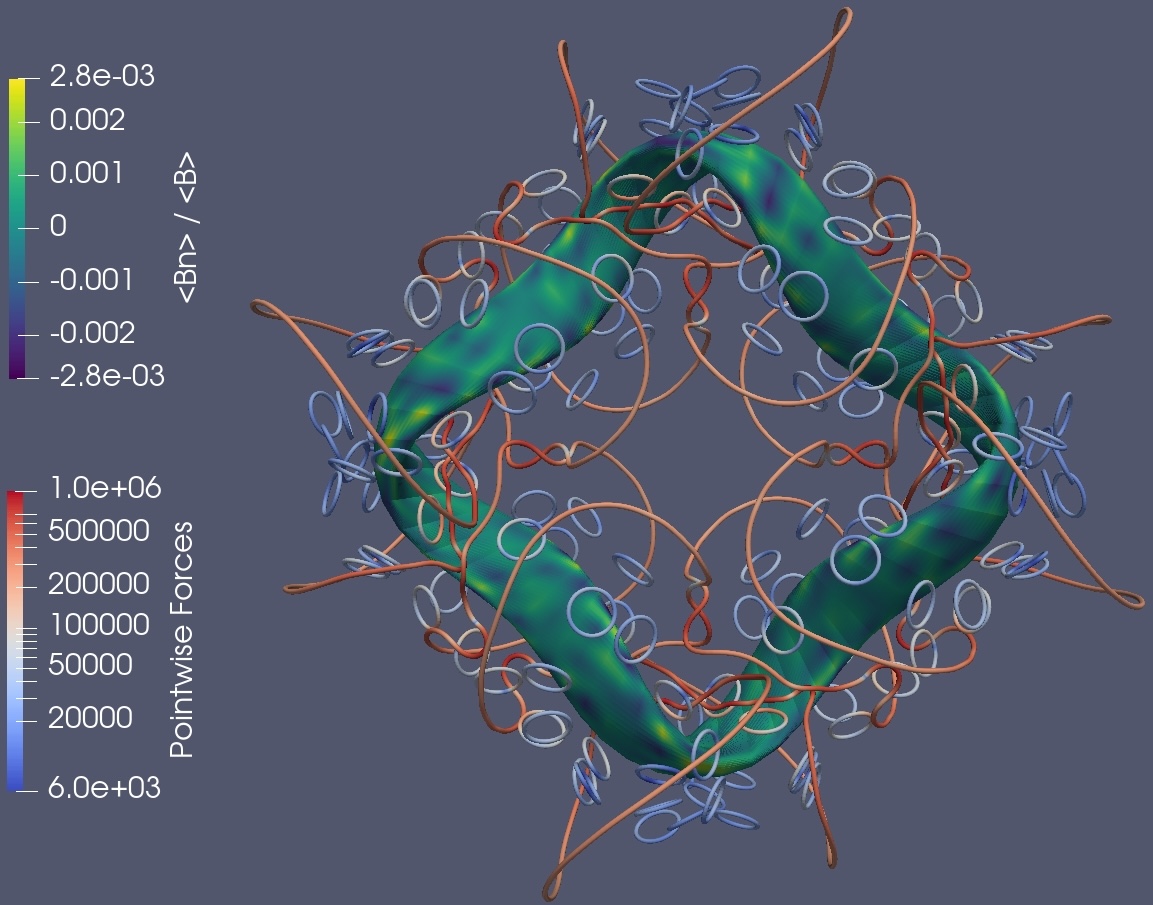}
    \caption{PSC array solution for the four-field-period Landreman-Paul QH stellarator.}
    \label{fig:QH}
\includegraphics[width=0.875\linewidth]{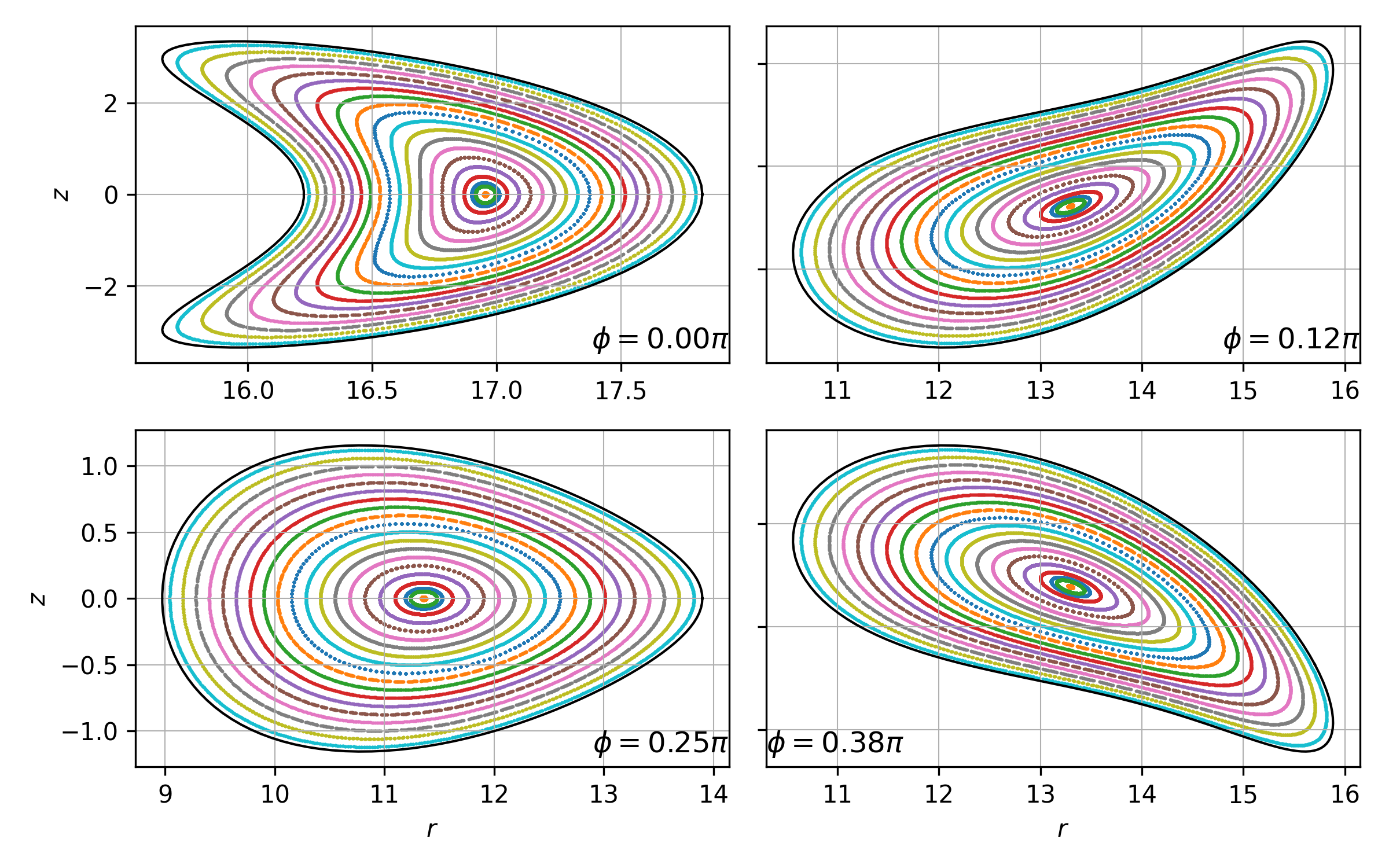}
    \caption{Poincaré plot for the Landreman-Paul QH coil design shows good magnetic surfaces.}
    \label{fig:QH_poincare}
\end{figure}

In contrast to the other examples, the Landreman-Paul QH stellarator is four-field period symmetric. QH stellarators can sometimes require quite complex modular or helical coils. The only optimized quasi-helical stellarator that has been built is HSX and it uses six modular coils per half-period~\cite{anderson1995helically,almagri1999helically}. Solutions for the reactor-scale Landreman-Paul QH stellarator have been found in Wiedman et al.~\cite{wiedman2023coil} using five modular coils per half-period and in Kaptanoglu et al.~\cite{kaptanoglu2024reactorscalestellaratorsforcetorque} using two modular coils per half-period alongside a dipole array. In the present work, we jointly optimize a PSC array with 19 coils per half-field-period all with fixed radius $1.23$m and two TF coils of total length $110$m and average curvature $\kappa \approx 1.45$m$^{-1}$. The accuracy of the solution is $\langle \bm B \cdot\hat{\bm n}\rangle/\langle B\rangle \approx 6.4\times 10^{-4}$. Empirically, we observed that reducing the $\langle \bm B \cdot\hat{\bm n}\rangle/\langle B\rangle$ errors substantially lower still obtained a degree of quasi-symmetry that was an order of magnitude worse than the original equilibrium; the degree of quasi-helical symmetry may be quite sensitive to field ripple from small coils. 
Therefore, we reduced the coil complexity at the expense of some quasi-symmetry breaking.  A minimum distance of $1$m is prescribed between TF coils, and the minimum distance between one PSC and a TF coil or two PSCs is $0.5$m.
We obtain a two-term average quasi-symmetry of $1.1\times 10^{-3}$, a value still substantially better than any stellarator in experimental operation. The optimized PSC currents vary between $0.83$-$2.5$ MA, the TF coil forces in Fig.~\ref{fig:QH} are roughly at the material tolerances, and Poincaré plots in Fig.~\ref{fig:QH_poincare} show good magnetic surfaces.

\subsection{Reactor-scale Schuett-Henneberg QA stellarator}\label{sec:henneberg_QA}
\begin{figure}
    \centering
    \includegraphics[width=0.92\linewidth]{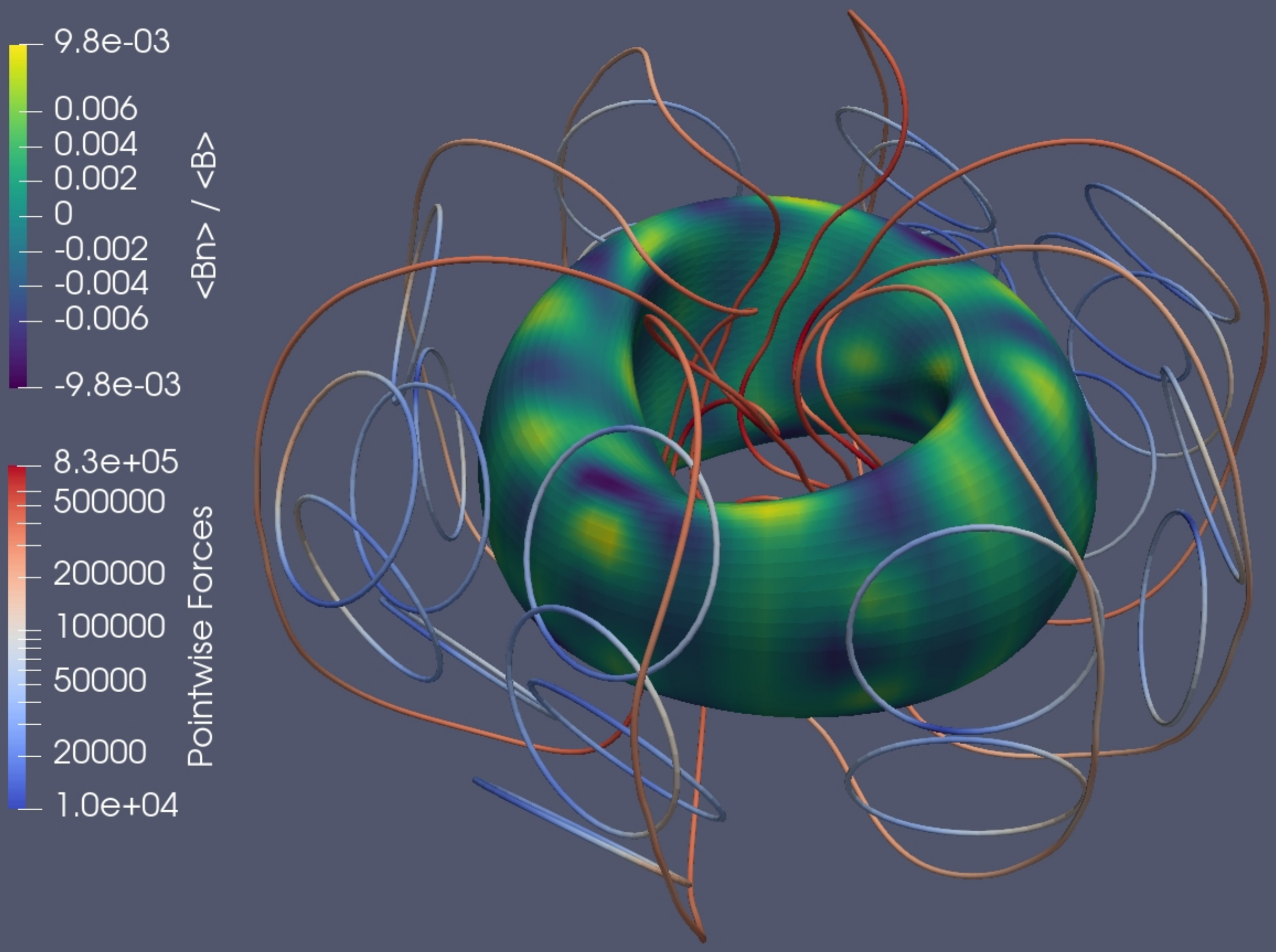}
    \caption{PSC array solution for the two-field-period QA Schuett-Henneberg stellarator.}
    \label{fig:henneberg}
    \includegraphics[width=0.875\linewidth]{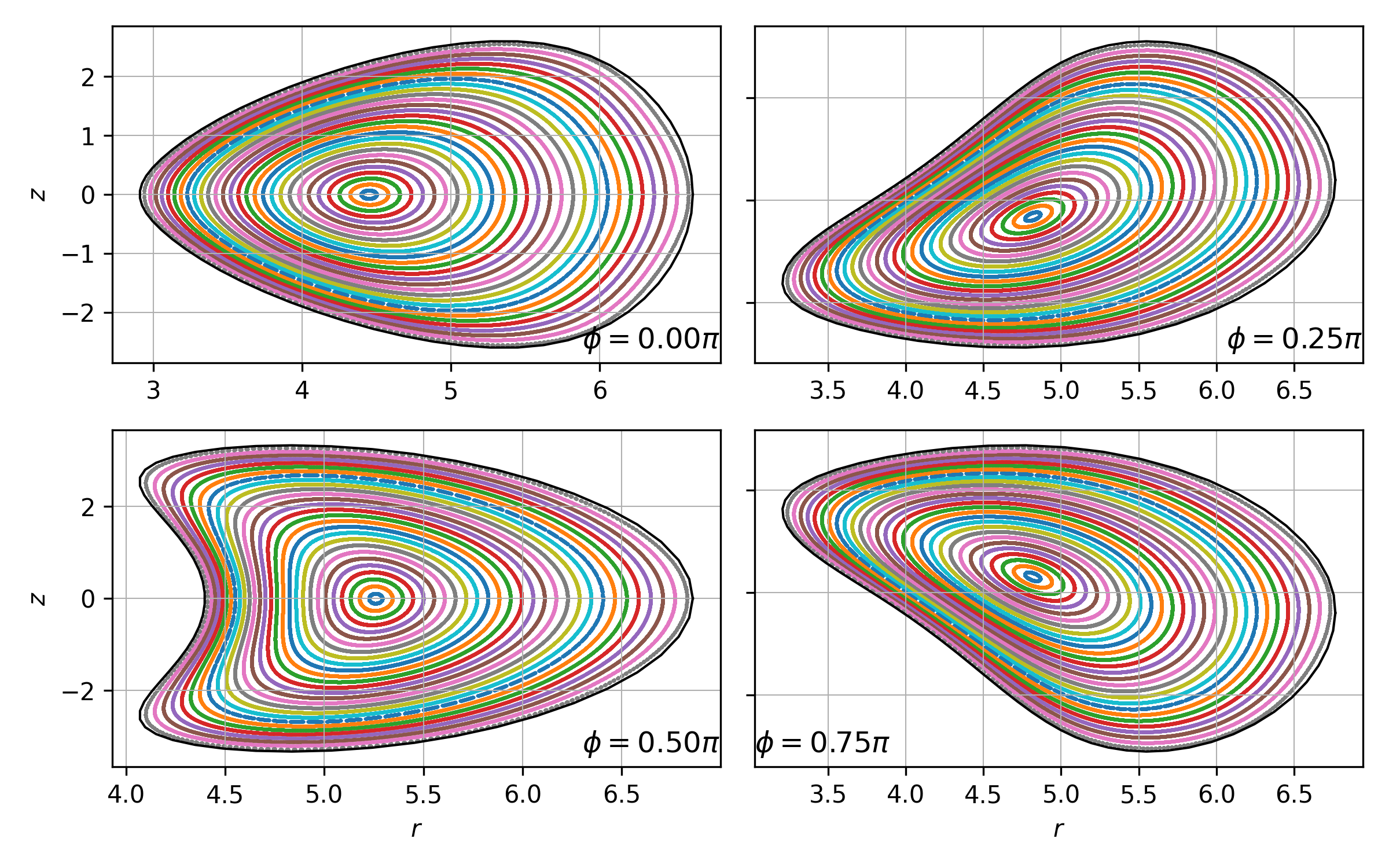}
    \caption{Poincaré plot for the Schuett-Henneberg coil design shows good magnetic surfaces.}
    \label{fig:henneberg_poincare}
\end{figure}

The two-field-period Schuett-Henneberg QA stellarator~\cite{schuett2024exploring} is a very compact device. Subsequently, it is difficult to place any PSCs on the inboard side of the plasma. The primary use is to mitigate the outboard modular coil ripple.
In Fig.~\ref{fig:henneberg}, we find an accurate solution with $\langle \bm B \cdot\hat{\bm n}\rangle/\langle B\rangle \approx 2.4\times 10^{-3}$ using an array of six PSCs per half-field-period with radius $3.26$m and two TF coils. The accuracy of the final solution is high enough to produce good Poincaré plots in Fig.~\ref{fig:henneberg_poincare} and the average normalized two-term quasi-axisymmetry error of approximately $3.9\times 10^{-3}$ is fairly close to the original value of $1.4\times 10^{-3}$.  Access to the plasma is very high for a reactor-design, especially from the top and bottom of the device where no PSCs are placed. The TF coils have total length of $81$m, average coil curvature $\kappa \approx 0.65$m$^{-1}$, and forces within the material tolerances. The PSC coil currents vary between 1.2-2.0 MA. During the course of optimization, we observed PSC coils placed in suboptimal locations, where we see order of magnitude smaller currents. These coils contribute very little to the solution and were removed after the optimization, followed by another round of optimization with the remaining coils. This is the well-known technique of sequentially-thresholded least-squares~\cite{Brunton2016pnas,kaptanoglu2023sparse, kaptanoglu2024topology} for promoting sparsity in the solution. 

\subsection{University-scale CSX stellarator}\label{sec:CSX}
Like permanent magnets, the use of passive coils may be a cost-effective way to build university-scale experiments. Moreover, such devices are well-positioned to study passive interaction and potential stabilization between PSCs and the plasma.
The Columbia Stellarator eXperiment (CSX) is currently being designed and aims to test predictions related to QA plasma behavior. 
The previous CNT experiment~\cite{pedersen2004columbia} was generated by a combination of two circular planar poloidal field (PF) coils
and two interlinked (IL) coils. Recent work in Baillod et al.~\cite{baillod2024integrating} investigated possible coil solutions for this new experiment and found that the degree of quasi-symmetry is highest when a set of windowpane coils are used to enhance shaping and experimental flexibility. The PF coils provide a simple background field throughout the vacuum vessel, so we removed these active window-pane coils and instead attempted a passive coil solution for the most quasi-axisymmetric plasma shape found in this recent work. 

\begin{figure}
    \centering
    \includegraphics[width=0.925\linewidth]{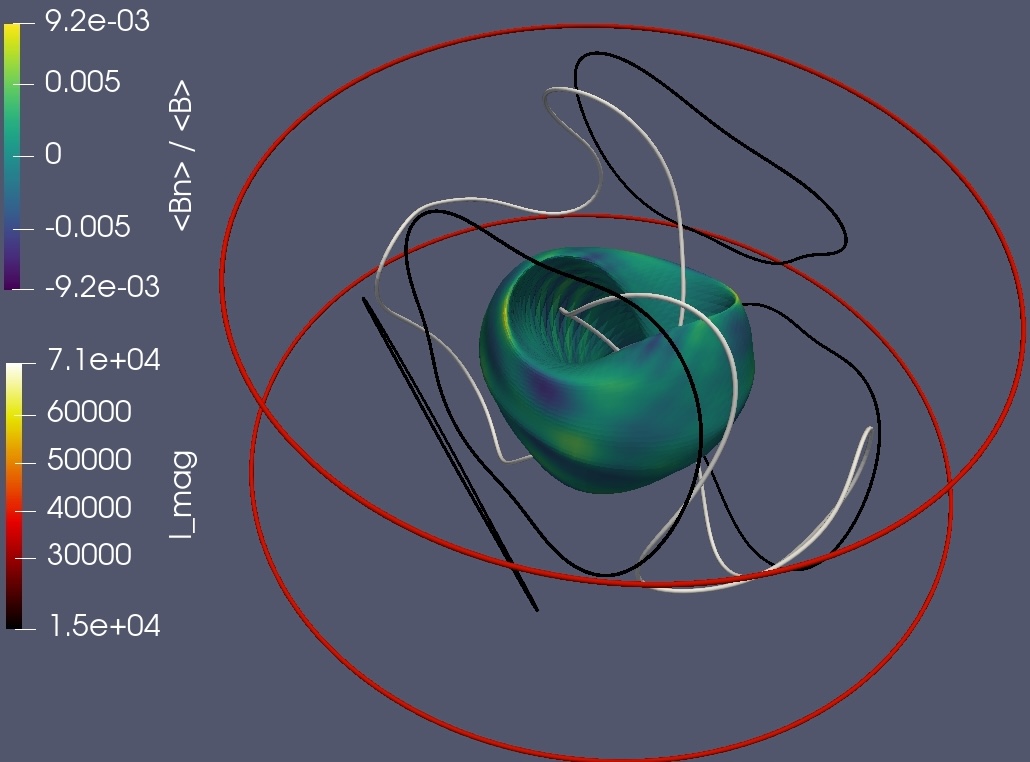}
    \caption{PSC array solution for the two-field-period QA CSX stellarator. The IL currents are colored white, the PF currents are red, and the PSC currents are black.}
\label{fig:CSX}\includegraphics[width=0.92\linewidth]{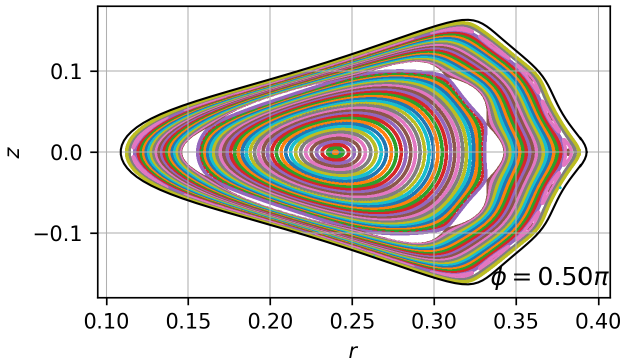}
    \caption{Poincaré plot for the CSX coil design shows good magnetic surfaces alongside a (2, 8) island chain.}
    \label{fig:CSX_poincare}
\end{figure}

We use only the PF and IL coils, with a single passive coil per half-field-period inside the vacuum vessel. The PF coils are fixed in shape and current, the IL coils are jointly optimized with the PSC, and the length of the IL coils are limited to $5.0$m as in the recent CSX design work. We let the passive coil shape vary, although it is constrained to be planar. We find that we can achieve a similarly accurate solution $\langle \bm B \cdot\hat{\bm n}\rangle/\langle B\rangle \approx 1.3\times 10^{-3}$ in Fig.~\ref{fig:CSX}, comparable with the two nonplanar window-panes per half-field-period solution in Baillod et al.~\cite{baillod2024integrating}. In addition the PSC by definition has no power supply, and has 15 kA of current, less than $1/3$ of the current that is driven in the window-panes. The length of the PSC curve is $2.89$ m and at closest is $4$ cm away from the IL coils. The IL coils are closest to the plasma with a minimum coil-plasma distance of $12$ cm. We find that optimizing the IL coils alone, with no PSC, produces a solution with over an order of magnitude larger errors. A Poincaré plot in Fig.~\ref{fig:CSX_poincare} illustrates good magnetic flux surfaces, although a (2, 8) magnetic island chain appears. 

\section{Conclusion}\label{sec:conclusion}
We have formulated and implemented the inverse magnetostatic optimization of passive superconductors. PSC arrays demonstrate significant promise as an opportunity to reduce the complexity of designing stellarator coils. We presented PSC array solutions for four different stellarator designs. 
Improved coil configurations, with or without passive coils, may be found in future work, as coil optimization problems generally exhibit numerous local minima and the tuning of the objective weights will depend on the tradeoff preferences of the user. There may be some stellarator designs that do not benefit for PSC arrays, as even reactor-scale passive coil currents are necessarily clamped at a few MA. For instance, an accurate PSC array solution for the Landreman-Paul QH stellarator was quite hard to find and the quasi-symmetry of the final design seemed quite sensitive to very small errors. In contrast, the Schuett-Henneberg QA design had average $\langle \bm B\cdot\hat{\bm n}\rangle / \langle B \rangle$ errors that were four times larger than the QH coil design, but the final design still had similar levels of quasi-symmetry to the original target plasma surface. 

 Furthermore, finding accurate solutions typically requires allowing the PSCs to move and orient in space. This would preclude a very simple structural support for the PSC array, such as a single toroidal shell. We also sometimes find that optimization performance improves when an initial optimization is performed in which only the passive coil centers can vary and the TF coils are fully varied. Then this optimization is further refined by additionally allowing the PSCs to rotate in space, and if accuracy allows, reducing the length of the TF coils. A similar strategy was useful when coil shape optimization was used.

The PSC array designs presented in this work almost certainly use more superconducting material than equivalently accurate designs using only modular coils. Plasma experimentalists must decide whether the degree of geometric simplicity and access to the plasma with PSC arrays may compensate for the increased costs of purchasing more material.

\section*{Acknowledgements}
This work was supported through a grant from the Simons Foundation under award 560651. 

\appendix

\section{Meissner effects in superconductors}\label{sec:meissner_discussion}
 Meissner effects are challenging to quantify for both Type I (mostly low-temperature) and Type II (mostly high-temperature) superconductors because it can depend strongly on the material properties, the thickness of the material, and so on. In this appendix, we include some discussion on these effects and an argument why it is subdominant for Type I superconductors that are small in size compared to the characteristic variation of the magnetic field. Quantifying Meissner effects is a challenge relevant to any superconducting coils, and therefore merits further investigation independent of the present work.

Classical, low-temperature (Type I) superconductors expel magnetic field, tending
to make $\mathbf{B}\approx0$ in the interior of the conductor. In
other words, they are almost perfectly diamagnetic. However, $\mathbf{B}$ is not exactly zero in the
material. First, $\mathbf{B}$ decays exponentially
in the interior of the conductor, where the scale length for the exponential
decrease is the London penetration length $\lambda$. The penetration depth can be on the order $10^{-7}$ or $10^{-6}$ m~\cite{zaitsev2002microwave}. 
%
In high-temperature superconducting (HTS) tapes used for fusion, usually the REBCO layer of the HTS tape is just $\sim 10^{-6}$ meters thick, while the tape itself might be $\sim 10^{-4}$ m~\cite{sundaram20162g}. Therefore, the width of the REBCO layer might be only slightly larger than the penetration length, and the veracity of approximating these currents as surface currents is unclear, even if these tapes were type-I superconductors with a true Meissner effect. 
%
Moreover, $\mathbf{B}$ is not reduced to zero completely
in Type II superconductors above a certain applied field $H_{c1}$
which is well below the critical field for superconductivity. 
In such superconductors, a mixed state with field penetration of the superconductor exists between the lower critical field $H_{c1}$ and the upper critical field $H_{c2}$. 
%
Some magnetic flux quanta penetrate the coil, and with time-varying fields, buildup of these fluxes can generate large internal forces. This is the case for all HTS-made coils and is independent of the passive nature of the coils considered here. 


\begin{figure}
    \centering
    \includegraphics[width=0.8\linewidth]{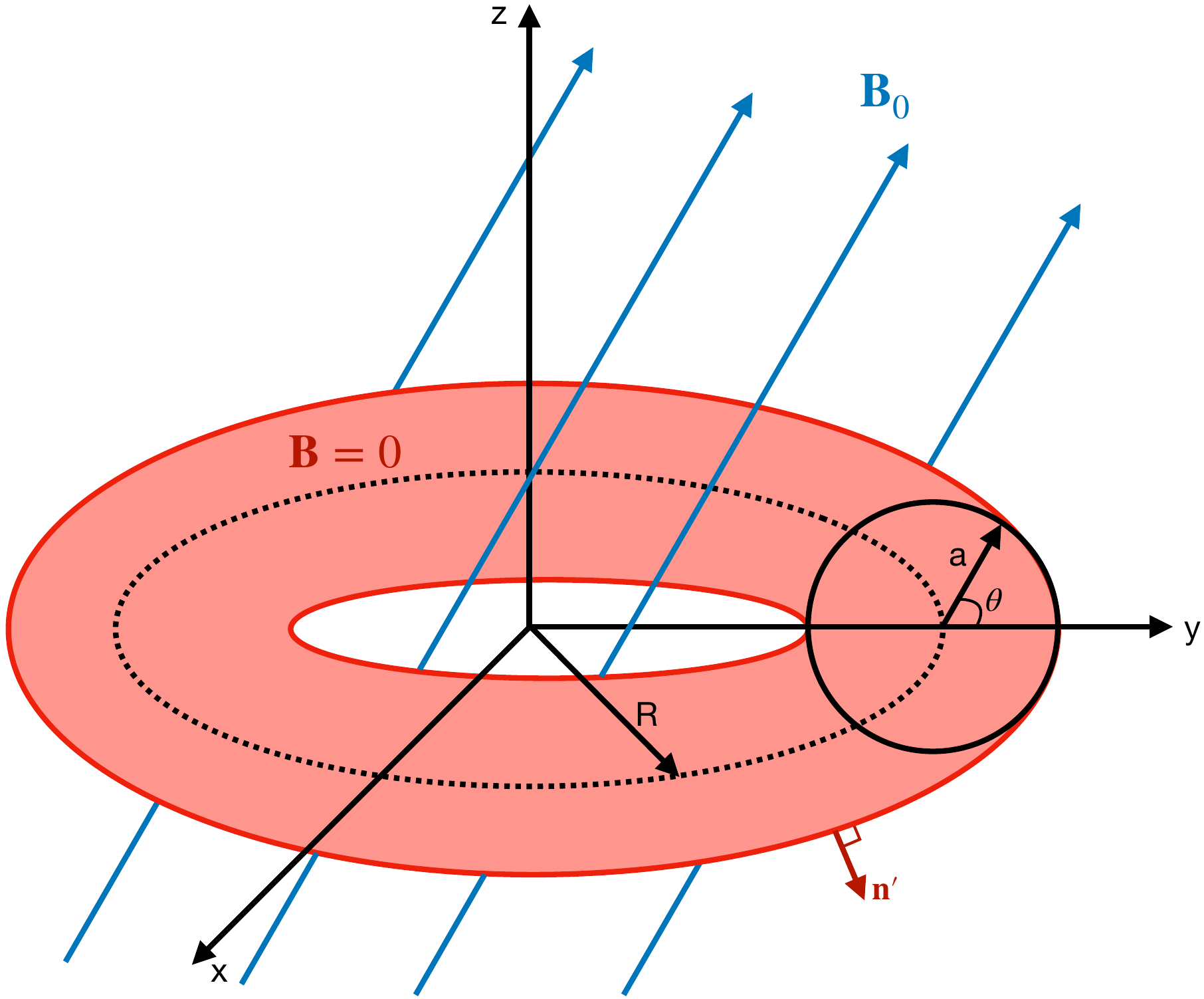}
    \caption{Geometry of a finite-cross section circular coil with a uniform $\bm B_0$ field on the scale $R$ of the coil. For clarity, the coil is visualized in a small aspect ratio.}
    \label{fig:uniform_field_calculation}
\end{figure}

It can be shown that the Meissner-related magnetic fields are subdominant in a circular Type-I superconducting coil with circular cross section. As earlier, denote the major radius of the coil as $R$ and the minor radius as $a$ with $R \gg a$. 

First, let us show that the induced currents associated with the flux-freezing and the currents associated with the Meissner effect are independent. For both properties, the associated currents must be surface currents; otherwise, the magnetic field would penetrate the interior of the conductor. Using the high aspect-ratio of the coil, we can zoom in on the coil so that it appears like an infinite straight cylinder. The current associated with flux-freezing moves parallel to the axis of the cylinder. Such a current does not create any magnetic field inside the cylinder by a simple application of Ampere's law. Thus, no extra shielding current from the Meissner effect is needed to make the flux-freezing associated magnetic field vanish inside the conductor.
%

In this case, a nontrivial Meissner effect appears only if there is an external applied field $\bm B_0$. We now show that the Meissner contribution to the magnetic field is identically zero if $\bm B_0$ is spatially uniform on the scale of the coil.
Consider a Type-I superconducting coil embedded in a spatially uniform but arbitrarily oriented $\bm B_0$ in Fig.~\ref{fig:uniform_field_calculation}, described in a simple toroidal coordinate system $(r', \theta, \phi)$ where,
\begin{align}
    \rho = R + r'\cos(\theta), \quad z =  r'\sin(\theta),
\end{align}
where $(\rho, \phi, z)$ is the cylindrical coordinate system centered in the coil, $r'$ is a minor radial coordinate, and $\theta$ is the poloidal angle. We model the coil as perfectly diagmagnetic, so there is a pure surface current. From electromagnetic boundary conditions, the surface current density is
$\bm J = \mu_0^{-1} \hat{\bm n}_0 \times \bm B_0$. In terms of Cartesian unit vectors,
\begin{align}
    \hat{\bm n}_0 = \cos(\phi)\cos(\theta)\hat{\bm x} + \sin(\phi)\cos(\theta)\hat{\bm y} + \sin(\theta)\hat{\bm z},
\end{align}
is the unit normal to the toroidal surface.
Consider a poloidal Amperian loop with tangent vector $\bar{\bm \gamma}$,
\begin{align}
    \int_0^{2\pi} \bm B_0\cdot\bar{\bm \gamma} d\theta = \mu_0\int_0^{2\pi}d\theta\int_0^b dr' \bm J\cdot\hat{\phi} = \mu_0 I_\text{enc}.
\end{align}
This integral vanishes since for $b < a$, $\bm J = 0$ and there is no enclosed current, $I_\text{enc} = 0$, by this loop. For $b > a$, $I_\text{enc} = 0$ still since, $\bm J \cdot\hat{\phi} = -\mu_0^{-1} \hat{\bm n}_0 \times \hat{\phi}\cdot\bm B_0$ and thus
\begin{align}
    \mu_0I_\text{enc} =-\bm B_0\cdot\int_0^{2\pi}d\theta\int_0^b dr' \bm n_0\times\hat{\phi}.
\end{align}
Each component of the vector integrand will integrate over $\theta$ to zero. Notice for $b > a$, the integral vanishes only because $\bm B_0$ is assumed spatially uniform. So there is no net current flow azimuthally in the coil. 
Similarly, consider a toroidal Amperian loop, 
\begin{align}
    \int_0^{2\pi} \bm B_0\cdot\bar{\bm \gamma} d\phi = \mu_0\int_0^{2\pi}d\phi\int_0^b \rho d\rho \bm J\cdot\hat{\theta} = \mu_0 I_\text{enc}.
\end{align}
For $b < R - a$, $\bm J = 0$ on the surface and $I_\text{enc} = 0$. For $b > R - a$, 
\begin{align}
\notag
\hat{\theta} &= -\sin(\theta)\cos(\phi)\hat{\bm x} - \sin(\theta)\sin(\phi)\hat{\bm y} + \cos(\theta)\hat{\bm z}, \\ 
\hat{\bm n}_0 \times\hat{\theta} &= -\hat{\phi},
\end{align}
and therefore the Amperian surface integral will
integrate to zero over $\phi$ in each component. 
To summarize, there is no net current flowing azimuthally or poloidally in the coil. 
In other words, if $\bm B_0$ is assumed constant on the scale of the coil, the Meissner effect provides no magnetic field contribution even for points immediately outside the coil. The only effect is to locally shield the magnetic field but otherwise does nothing to modify the total field outside the coil.

To arrive at a nontrivial Meissner contribution to the magnetic field, $\bm B_0(\bm x)$ must vary spatially. We can expand $\bm B_0(\bm x)$ at $\bm R = R[\cos(\phi), \sin(\phi), 0]$, and assume the scale $L_0$ of magnetic field variation is large, $L_0 \gg R$. Then to next leading order there is a magnetic field contribution $\sim (\bm x^T - \bm R^T)\cdot\bm B_0(\bm R) / L_0$ which can now produce net currents but scales as $R/L_0 \ll 1$.
\bibliography{passive_superconducting_coils}
\end{document}